\documentstyle[12pt,fleqn]{article} 
\Huge
\newcommand{\beq}{\begin{equation}}
\newcommand{\eeq}{\end{equation}}
\newcommand{\bea}{\begin{eqnarray}}
\newcommand{\eea}{\end{eqnarray}}
\newcommand{\beau}{\begin{eqnarray*}}
\newcommand{\eeau}{\end{eqnarray*}}

\newcommand{\eqn}{\begin{equation}}
\newcommand{\eqnend}{\end{equation}}

\def\sp{\kern 1em}

\begin{document}
\title{\bf Self-consistent current-voltage characteristics of
superconducting nano-structures.}
\author
{{\bf A. Martin and C.J. Lambert}\\
{\bf School of Physics and Materials},\\
{\bf Lancaster University, Lancaster, LA1 4YB, England}}
\maketitle
\noindent
\begin{abstract}

 By solving the Bogoliubov - de Gennes
     equation self-consistently in the presence of a non-equilibrium
     quasi-particle distribution, we compute
     the current-voltage characteristic of a phase coherent
     superconducting island
     with a tunnel barrier at one end.
     The results show significant structure, arising from the
     competition between scattering processes at the boundaries of the
     island and modification of the order parameter by quasi-particles and
     superflow. This structure is not present in
     non self-consistent descriptions of normal-superconducting
     nano-structures.

\end{abstract}

\vskip 0.5cm
\noindent
{\bf PACS numbers: 71.55.Jv, 71.30.+h, 05.40.+j}

\newpage\clearpage\

It is now well established that coherent Andreev scattering provides the
key to understanding transport in mesoscopic superconductors and
normal-superconducting (N-S) interfaces. For example zero bias anomalies[1]
can be understood through a description based
on multiple Andreev scattering in N-S tunnel junctions[2-5],
while phase periodic
conductances in N-S-N nano-structures[6]
are understandable through theories of coherent transport which neglect
inelastic scattering[7-14].
All of the above theoretical descriptions are based on non self-consistent
solutions of the Bogoliubov de-Gennes  equation or corresponding
quasi-classical equations and are not capable of describing the modification
of a superconducting order parameter by a transport current.
Effects of this kind are observable as super-gap structure in the differential
conductance of N-S tunnel junctions and point contacts[15,16], but to date,
there exists no quantitative theoretical framework for their understanding.
For structures smaller than the inelastic phase breaking length $l_\phi$,
such features cannot be ascribed to quasi-particle \lq\lq heating ", because
the energy of the electron is preserved during its passage through the sample.
Instead, any modification is a hot electron effect and requires a description
which takes into account the non-equilibrium distribution of electrons
within the sample.

A theoretical description which encompasses both
non-equilibrium effects of this kind and phenomena associated with phase
coherent transport does not
currently exist. The aim of this Paper is to provide the first
self-consistent description of phase coherent transport in a N-S-N
structure, based on exact solution of the BdG equation.
Motivated by the success of the Blonder,
Tinkham and Klapwijk[17] (BTK)
calculation for the current-voltage (I-V) characteristic of a N-S
interface with a delta function scatterer, we examine the simplest
possible generalisation which is capable of highlighting the new
physics which emerges from a self-consistent description.
The system of interest is a mesoscopic scattering region containing a
superconducting island
connected to perfect normal leads, which are in turn connected to external
reservoirs at chemical potentials $\mu_1$ and $\mu_2$.
The system length $L$ is assumed to be smaller than a
quasi-particle phase breaking length and therefore a description, which
incorporates quasi-particle phase coherence throughout the system
is appropriate. The main question of interest is whether or not
such a description yields observable super-gap structure, which is absent from
a BTK desciption, thereby obviating the need to introduce ad hoc
 heating effects.

Before discussing quantitative results, it is useful to identify
characteristic voltages which are missing from the  BTK description[17].
For convenience, consider the zero temperature limit
and choose $\mu_1\ge\mu_2$. In this case electrons (holes) are incident
on the island from the left (right) reservoir  over an energy interval
$0<E<\mu_1-\mu$ ($0<E<\mu-\mu_2$), where $\mu$ is the self-consistently
determined condensate chemical
potential. The associated current will both suppress the magnitude
of the order parameter and generate a
phase gradient.
For a homogeneous superconductor with an order parameter $\Delta(x)=
\Delta_0{\exp}[iv_sx]$, the energy gap for excitations parallel (anti-parallel)
to the phase gradient $v_s$ is
$\mu_-=\Delta_0 - p_Fv_s$ ($\mu_+=\Delta_0+p_Fv_s$), where $p_F$ is the Fermi
momentum. For a long enough island, where the order parameter at the  centre
of the island is approximated by the above form,
excitations incident on the superconductor with energies less than
these values will be reflected and therefore in addition to the
voltage $\Delta_0/e$, one might naively expect
the I-V
characteristic to show some feature when the reservoir potentials
satisfy $\mu_1-\mu=\mu_\pm$ or $\mu-\mu_2=\mu_\pm$.
 In addition one might expect features to occur for those values of
$\mu_1-\mu_2$ at which $\Delta_0-p_Fv_s=0$
and
at which the self-consistent value of $\Delta(x)$ vanishes everywhere.

To obtain a self-consistent description,
we solve the Bogoliubov - de Gennes equation

\begin{equation}
{ H}(x)\left(\matrix{u_n(x) \cr
v_n(x)} \right )
= E_n\left(\matrix{u_n(x) \cr
v_n(x)} \right )
\end{equation}
%
with a Hamiltonian

\begin{equation}
{H(x)} =
\left(\matrix{[-(\hbar^2/2m)\partial_x^2+u(x) -\mu]
& \Delta (x) \cr
\Delta^*(x)
& -[-(\hbar^2/2m)\partial_x^2+u(x) -\mu]
\,}\right)
\end{equation}

\noindent
where  $\mu$ is the condensate chemical
potential, $u(x)$ is the normal scattering potential and $\Delta(x)$
the superconducting order parameter,
defined self-consistently by the following equation
\begin{equation}
\Delta(x)=V(x)\left(
{\sum_{E_n>0}}'u^*_n(x)v_n(x)-{\sum_{{E_n>0}\atop{\sigma}}}'u^*_n(x)v_n(x)
\langle \langle\gamma_{n\sigma}^{\dagger}(x)\gamma_{n\sigma}(x)\rangle\rangle
\right)
\end{equation}

\noindent
In this expression, primes on the sums indicate that only terms with $E_n$
less than some cut-off $E_c$ are to
be included, due to the fact that the electron-electron interaction is only
attractive over a small range of energies near the Fermi surface,
$\gamma_{n\sigma}^\dagger$ creates a Bogoliubov quasi-particle
and double angular brackets indicate a trace
over the density matrix of the system.
In what follows, the pairing potential $V(x)$ is chosen to equal a
constant for $0<x<L$ and to vanish outside this interval. The normal scattering
potential is chosen to be $u(x)/\mu_0=(2Z/k_F)\delta(x)$, where
$\mu_0$ is the condensate chemical potential in the absence of
an applied voltage and
$k_F=(2m\mu_0/\hbar^2)^{1/2}$. For a given choice of $L, Z, E_c, V_0$ and
reservoir  potentials, both the magnitude and phase of $\Delta(x)$
will be computed at all points in space, along with the condensate
chemical potential $\mu$.

Since we are interested in an open system,  equation (3) involves
sums over all incoming scattering states, integrated over all
$E < E_c$. At zero temperature, for the case $\mu_1>\mu>\mu_2$, quasi-particle
states corresponding to incoming electrons (holes) are incident from
reservoir 1 (2) over energy intervals $\mu_1-\mu$ $(\mu-\mu_2)$. Assuming these
intervals are less than the cut-off $E_c$ and if a scattering state of energy
$E$ corresponding to an incident quasi-particle of type $\alpha$ from reservoir
$i$ has a particle (hole) amplitude $u_{i\alpha}(x,E)$ ($v_{i\alpha}(x,E)$),
then equation (3) reduces to

\begin{eqnarray}
\Delta(x)
&=& V(x)\sum_{i=1}^2 \frac{1}{2}
\int_0^{E_c} ((u_{i -}^*(x,E)v_{i -}(x,E))
+ ((u_{i +}^*(x,E)v_{i +}(x,E)))dE \nonumber \\
&-& V(x)\int_0^{\mu_1-\mu} (u_{1 +}^*(x,E)v_{1 +}(x,E))dE \nonumber \\
&-& V(x)\int_0^{\mu-\mu_2} (u_{2 -}^*(x,E)v_{2 -}(x,E))dE
\end{eqnarray}

To calculate scattering solutions in the region occupied by the island,
we start from an initial guess for $\Delta(x)$ and $\mu$ and
divide the interval $0<x<L$ into a large number of small
 cells of size $<<k_F^{-1}$, within
which $\Delta(x)$ and $ u(x)$ are assumed constant.
If $T(x_0)$ is the matrix obtained by producting together transfer
matrices associated with all cells in the interval $0<x<x_0$
and then as outlined in appendix 1 of reference[18],
 the scattering matrix $S$ of the island can be obtained from
 the transfer matrix $T(L)$.
Within the external leads,  the most general eigenstate
of ${ H}$ belonging to eigen-energy E is a linear superposition of
plane waves.
For a given incoming plane wave, a knowledge of $S$ yields the plane wave
amplitudes on the left side of the island, which can be combined with
$T(x_0)$ to yield the wavefunction at $x=x_0$.
Given these solutions, $\Delta(x)$ is re-evaluated using equation (4)
and a new choice for $\mu$ is obtained by insisting that
the currents $j_1$ and $j_2$ in the leads attached to
reservoirs 1 and 2 are equal. This process is repeated until
the root mean square difference between successive
order parameters is less than $1\%$ of the magnitude of $\Delta(L/2)$.

In what follows all results are for
an island of length $Lk_F=750$, a cut-off
of $E_c=0.085\mu_0$ and a pairing potential of magnitude $V=0.28\mu_0$.
For an infinite homogeneous superconductor with no current flowing
and a density of states $n(0)$,
 BCS theory predicts a bulk order
parameter of magnitude $\Delta_0=E_c/{\rm sinh}\{1/[n(0)V]\}$,
which for this
choice of parameters yields $\Delta_0\simeq 0.005\mu_0$. As an example of the
results obtained, for an island
with no barrier (ie $Z=0$), figure 1 shows self-consistent
results for the magnitude $\vert\Delta(x)\vert$ and phase $\phi(x)$
of the order parameter, for various applied reservoirs potential differences.
As expected, $\vert\Delta(x)\vert$ reaches a maximum value
at $x=L/2$ and is suppressed at the ends of the
island, on a length scale $\xi=k_F^{-1}\mu/\vert\Delta(L/2)\vert$,
whereas the corresponding phase gradient
$v_s(x)=\nabla\phi(x)$
is almost a constant.
Furthermore the zero voltage value of $\Delta(L/2)$, (denoted $\Delta_0$ in
what follows)  agrees with the BCS prediction.
In what follows, we denote $v_s(L/2)$ and
$\vert\Delta(L/2)\vert$ by $v_s$ and $\Delta_s$ respectively.

By repeating these calculations for a range of reservoir potentials and barrier
strengths, one obtains I-V curves, whose derivative yields the
differential conductance shown in figure 2. Clearly these curves exhibit
structure which is not present in a non self-consistent description[17]. To
identify the underlying physical processes, figure 3
shows self-consistently determined values of $\Delta_s$ and the
various characteristic voltages identified above, plotted against the
reservoir potential difference.
For each value of $Z$, the upper and lower dashed lines of figure 3
show results for $\mu_+$ and
$\mu_-$ respectively, while the thin solid line shows $\Delta_s$.
The upper and lower thick solid lines show
values of $\mu_1-\mu$ and $\mu-\mu_2$ respectively. For $Z=0$, where
$\mu=(\mu_1-\mu_2)/2$, the latter are equal. More generally, for the
range of voltages studied, one finds $\mu =\mu_2 + \alpha(\mu_1-\mu_2)$, where
$\alpha=0.5, 0.421, 0.241, 0.126$ for $Z= 0, 0.25, 0.55, 0.83$, respectively.
Maxima and minima in the differential conductance of figure 2
are associated with various crossings in figure 3.

Consider for example the $Z=0.55$ and $Z=0.83$ results, where the first maximum
in $G_{N}^{-1}\partial I/\partial(\mu_1-\mu_2)$ corresponds to the crossing
$\mu_1-\mu=\Delta_s$. For these structures, provided
$\mu-\mu_2<\mu_-$, excitations from the right reservoir (2)
 are almost completely
Andreev reflected at $x=L$
and therefore the conductance is dominated by scattering
of electrons from the left reservoir at $x=0$. In this limit, it is of interest
to examine the quantity
 $G_{N}^{-1}\partial I/\partial(\mu_1-\mu)$, which in the absence of
quasi-particle
 transmission, is equivalent to the left boundary conductance, examined by
 BTK. The solid lines of figure 4 show self-consistent results  for
 this quantity. For comparison, the dashed lines show non self-consistent
 results (which
 are essentially those of BTK[17])
 obtained by insisting that $\mu=\mu_2$ and  for $0<x<L$,
$\Delta(x)=\Delta_0$
For $Z=0.83$ the dashed and solid curves of figure 4 are in good agreement,
reflecting the fact that for large $Z$ the current is small
and therefore $\Delta(x)$ is not significantly modified.
For the smaller barrier strength, the self-consistent conductance differs
significantly from the BTK curve. In the presence of quasi-particle
transmission [18,20], the resistance of a N-S-N structure does not reduce to
the sum of two boundary resistances. Consequently even for $Z=0.83$, the
N-S-N differential conductance of figure 2
shows extra structure which is absent from a boundary conductance calculation.
For example the second peak of the $Z=0.83$ results of figure 2
corresponds to the crossing
$\mu -\mu_2=\Delta_s-v_sp_F$,
while the minimum between these peaks corresponds to the crossing
$\mu_1-\mu=\Delta_s+v_sp_F$.
For $Z=0.55$ the peaks at $\mu_1-\mu$ and $\mu-\mu_2$ are no longer
separated, but again a minimum occurs at $\mu_1-\mu=\Delta_s+v_sp_F$.
For this value of $Z$, a maximum occurs at $\mu_1-\mu_2 \sim 3\Delta_s$,
at which the magnitude of $\Delta(L/2)$ starts to become significantly
reduced by the current.

To obtain the above results,
we have presented the first self-consistent description
of a superconducting nano-structure, which incorporates quasi-particle
phase coherence and non-equilibrium effects.
The computed current-voltage characteristic of a
single superconducting island is the result of  several competing
phenomena associated with quasi-particle scattering from boundaries
and modification of the order parameter by both superflow
and a non-equilibrium quasi-particle distribution. This competition
produces significant new structure, particularly at energies above
$\Delta_0$, which is not contained within
a non self-consistent description.
This structure is observable experimentally [19]
but in earlier discussions[15] has been dismissed as a distortion
 due to heating.
\vskip2.0cm
\noindent
{\Large\bf Note added in proof}

A similar calculation has been carried out recently by F.Sols and
J. Sanchez-Canizarez, in which the superconductor is treated as an incoherent
reservoir[21]. Where agreement is expected, their results are comparable
with those reported here.

\noindent
{\Large\bf Acknowledgments}

 This work was supported in part from EEC via an HCM grant,
 by the EPSRC, NATO, and the Institute for Scientific Exchange.
We thank F.Sols, J. Sanchez-Canizarez, S.J. Robinson and V.C. Hui
for interesting discussions.

\newpage\clearpage
{\Large\bf Figure captions}
\\ \\
{\bf Figure 1.}
Results for the case $Z=0$.
The upper figure shows the form of the attractive interaction
$V(x)$ (in units of $\mu_0$) used in all the calculations of this paper.
The middle and lower graphs
show the self-consistent forms the magnitude and phase of
$\Delta(x)=\vert\Delta(x)\vert\exp i\phi(x)$. The results with the largest
values of $\vert\Delta(x)\vert$ and constant phase correspond to
 $\mu_1-\mu_2 = 0$. In order of decreasing $\vert\Delta(x)\vert$ and increasing
 phase gradient,  remaining results correspond to  $\mu_1-\mu_2 = 0.006$
 and  $\mu_1-\mu_2 = 0.01$, respectively.
\\ \\
{\bf Figure 2.}
Self-consistent results for the differential conductance of a superconducting
dots, with a delta function barrier of strength $Z$, located at $x=0$.
\\ \\
{\bf Figure 3.}
For each value of $Z$, the upper (lower) dashed line shows self-consistent
results for the voltages $\mu_+=\Delta_s+p_Fv_s$
($\mu_-=\Delta_s-p_Fv_s$), while the thin solid line shows $\Delta_s=
\vert\Delta(L/2)\vert$. The upper (lower) thick solid line shows
self-consistent values of $\mu_1-\mu$ ($\mu-\mu_2$). For $Z=0$,
the latter are equal.
\\ \\
{\bf Figure 4.}
The solid lines show self-consistent results for $G_{N}^{-1}\partial I/\partial
(\mu_1-\mu)$. The dashed lines show non self-consistent results
obtained by setting $\Delta(x)=\Delta_s$ for $0<x<L$ and zero elsewhere.
\\ \\
\end{document}